# Coexistence of monoclinic and tetragonal phases in PMN-PT single crystal

E. B. Araújo, C. A. Guarany, K. Yukimitu, and J. C. S. Moraes

Universidade Estadual Paulista, Departamento de Física e Química, Caixa Postal 31

15385-000 Ilha Solteira, SP - Brazil

A. C. Hernandes

Universidade de São Paulo, Instituto de Física de São Carlos, Caixa Postal 676

13565-670 São Carlos, SP – Brazil

The purpose of the present work is to report studies on structural phase transition for PMN-xPT single crystal, with melt $PbTiO_3$ composition around the MPB (x = 0.35 mol %), using infrared spectroscopy technique. The study was centered on monitoring the behavior of the $\nu_1$-(Nb-O), $\nu_1$-(Ti-O) and $\nu_1$-(Mg-O) stretching modes as a function of temperature. Singularities observed between 198 and 300 K suggested a possible coexistence between the monoclinic and tetragonal phases, not revealed by others techniques, for PMN-PT composition studied near MPB.

## Introduction

Since the discovery of a new monoclinic phase in ceramic solid solutions $Pb(Zr,Ti)O_3$ (PZT) in 1999,[1] several works were published to understand the mechanisms responsible for high piezoelectric response of these materials. This monoclinic phase was observed for PZT compositions near the *morphotropic phase boundary* (MPB), initially interpreted as the coexistence of the rhombohedral ($F_R$) and tetragonal ($F_T$) ferroelectric phases, at room temperature in the composition-temperature phase diagram.[2] Based on subsequent studies, supported by different techniques such as high-resolution synchrotron x-ray powder diffraction,[3] [4] [5] Raman spectroscopy,[6] [7] dielectric measurements [8] and theoretical electromechanical studies,[9] the MPB was redefined assuming the monoclinic phase as responsible for high piezoelectric response in PZT ceramics.[10]

Within a short time interval, another monoclinic phase was also observed in similar solid solution $(1-x)Pb(Mg_{1/3}Nb_{2/3})O_3$-$xPbTiO_3$ (PMN-PT),[11] which also exhibits an analogous MPB between the rhombohedral and tetragonal ferroelectric phases in the PMN-PT composition-temperature phase diagram as observed in PZT. The PMN-PT is known for its exceptional electromechanical properties, sometimes one order of magnitude larger than classical PZT ceramics.[12] Subsequent to these studies, the monoclinic structure was able to be considered as a natural link between the tetragonal and rhombohedral phases in PZT or PMN-PT phase diagram that remain under intense investigations. An interesting review on advances to understand the role of the monoclinic phases on structure and high piezoelectricity in lead oxide solid solutions was published in 2002.[13]

The comprehension of the monoclinic phase in systems like PZT and PMN-PT near MPB is complicated and requires systematic studies in order to be correctly understood. In this context, spectroscopic techniques, such as Raman and infrared spectroscopy (IR), are useful tools to investigate phase transitions in polycrystalline materials and single crystals. Infrared spectroscopy can be used to study ferroelectric phase transitions since infrared vibrational frequencies, and consequently the interatomic forces, are affected by the onset of the ferroelectric state due to temperature phase transitions. Recently, we used the IR technique to study the *monoclinic* → *tetragonal* phase transition at PZT compositions system near the MPB to better understand the role of the monoclinic phase in this system.[14] [15]

In the literature, there are relatively few works reporting structural phase transition studies in PMN-PT single crystals using Raman or infrared spectroscopy, in the last two years there were only two works published. The first one was an extensive study based on dielectric and Raman spectroscopy measurements using a complete PMN-PT set of samples.[16] The second one was an interesting work involving infrared and Raman spectroscopy centered on lattice dynamic studies of the cubic, tetragonal and rhombohedral phases of PMN-PT single crystal. However, this last study was conducted for PMN-PT composition out of the MPB (PMN-29%PT).[17]

The purpose of the present work is to report a structural phase transition study on PMN-PT single crystal, with composition around the MPB, using infrared spectroscopy technique. Consequently, the *monoclinic* → *tetragonal* phase transition in this crystal were studied and compared with previous results obtained from other researchers. As a result, this work suggested the coexistence between monoclinic and tetragonal phases for a specific PMN-PT composition around MPB ($x = 0.35$), which previously were not revealed by other techniques.

## Experimental

In the present study, $(1\text{-}x)Pb(Mg_{1/3}Nb_{2/3})O_3\text{-}xPbTiO_3$ (PMN-PT) samples of crystals grown by the Bridgman technique, with x = 0.35 mol % PT melt composition, were used. More details regarding the crystal growth process are described elsewhere.[18]

An infrared measurement requires a small amount of powdered sample (~20 mg) with uniform grain size distribution. Using PMN-PT crystal powders dispersed in pressed KBr discs, infrared measurements were performed using a Nicolet Nexus 870 FT-IR in the wavenumbers range 1000-400 $cm^{-1}$ (each IR spectrum was recorded with 2 $cm^{-1}$ resolution). The KBr pellet was mounted inside a cryostat with KBr windows attached on the spectrometer and each spectrum was recorded in isotherms at temperatures between 100 and 470 K. The temperature of the cryostat was controlled manually within ±2 K accuracy. To certify the homogeneity in temperature of the KBr pellet, and consequently the PMN-PT temperature, each spectrum was collected after 1 hour under stable temperature (isotherm regime).

## Spectra fitting procedure

In order to analyze the infrared spectra as a function of temperature, each spectrum recorded in isotherm regime was fitted individually using the computational least-square method. The IR absorption band is associated to $BO_6$ octahedra. To apply the computational fitting, vibrations of the $BO_6$ octahedra were treated as simple harmonic oscillators. Under these considerations, each absorption mode was represented by a lorentzian function:

$$I = I_0 \frac{AW}{(\omega - \omega_f)^2 + W^2} \tag{1}$$

where $I_0$ is the constant, $\omega$ the experimental frequency, $\omega_f$ the fitting frequency, $A$ is the area under curve (associated with the number of the oscillators) and $W$ is the half width.

The number of lorentzian functions used to fit each spectrum was assigned based on possible vibration modes in PMN-PT as will be discussed afterwards. The PMN-PT presents a general $ABO_3$ perovskite structure and different phases depending on PMN/PT ratio. The infrared vibrations for this perovskite family may be explained based on vibrations of the $BO_6$ octahedron (B = Mg, Nb and Ti), similarly to classical perovskites such as $BaTiO_3$, $SrTiO_3$, $PbTiO_3$ and $PbZrO_3$ octahedron.[19 20 21] In these structures, $BO_6$ octahedron presents four distinct vibration modes: $\nu_1$-*stretching* at higher frequency, and lower frequencies $\nu_2$-*torsion*, $\nu_3$-*bending* and $\nu_4$-cation-($BO_3$) vibrations.[21] In the present investigation, only the $\nu_1$-*stretching* vibration mode will be studied, considering that $\nu_2$, $\nu_3$ and $\nu_4$ bands occur below the available experimental frequency range (4000-400 $cm^{-1}$) used in this work and for this reason will not be considered in our discussion.

In order to clarify the behavior of vibration modes as a function of the temperature, we decomposed each IR spectrum assuming at least three lorentzian curves, individually associated to $\nu_1$-(Nb-O), $\nu_1$-(Ti-O) and $\nu_1$-(Mg-O) stretching modes in the $BO_6$ octahedron. In other words, an observed infrared spectrum for PMN-PT is in fact a composition from these modes. Nevertheless, it is important to note that lorentzian numbers were not arbitrarily attributed but were physically linked with the expected band structure for PMN-PT. In principle, the assumption of three lorentzian curves must be considered at high symmetry (cubic structure) while at low symmetry (monoclinic structure) an additional lorentzian curve must be included in fitting. However, in this work three lorentzian curves were used over the entire temperature range because $\nu_1$-(B-O) modes are broad.

## Results and discussion

Several infrared measurements were recorded in the 100-400 K temperature range destined to study the structural phase transition in PMN-35PT. Fig. 1 shows infrared spectra from 900 $cm^{-1}$ to 400 $cm^{-1}$ for selected temperatures. In this frequency interval, a broad band was observed for each spectrum from 870 $cm^{-1}$ to 472 $cm^{-1}$, which presents a maximum absorbance and a shoulder around 617 $cm^{-1}$ and 531 $cm^{-1}$. This band is associated to $\nu_1$-$NbO_3$, $\nu_1$-$TiO_3$ and $\nu_1$-$MgO_3$ stretching modes in PMN-PT structure. As we can see, the curve shape slightly changes when temperature increases. At high temperature (470 K), the shoulder tends to disappear, the maximum absorbance shifts to 590 $cm^{-1}$ and consequently the spectrum assumes a more symmetric shape. These discrete changes as a function of the temperature may be attributed to the phase transitions involved in this system, introduced by distortions in $BO_6$ octahedra. Recently, other researchers have also observed similar behavior for PMN-29PT.[17] In the present work, these small changes on spectra shape were resolved by a detailed analysis from computational fittings.

Fig 2 shows graphically how fits were performed using three lorentzian functions, associated to three isolated $NbO_6$, $TiO_6$ and $MgO_6$ units, which reproduces the experimental infrared spectra. Computational fits using four and six lorentzian functions were also performed but will not be discussed in this work in details, these results did not present physical meaning. As $BO_6$ octahedra presents a broad absorption band on infrared, it is accepted that these bands can be represented satisfactory by three lorentzian functions.

Fig. 3 shows the $\nu_1$-(Nb-O), $\nu_1$-(Ti-O) and $\nu_1$-(Mg-O) stretching modes behavior as a function of the temperature. As we can see, all modes were sensible to changes as a function of the temperature. The $\nu_1$-(Nb-O) mode approximately decreases from 577 $cm^{-1}$ to 563 $cm^{-1}$ when the temperature increases from 100 K to 400 K. This behavior can be explained by an inverse relationship between atomic separation and vibrational frequency if there is no structural phase transition in the observed temperature range. However, in the same temperature interval, the $\nu_1$-(Ti-O) and $\nu_1$-(Mg-O) modes exhibits a distinct behavior, in principle, if compared with $\nu_1$-(Nb-O) mode. As observed, when temperature increases from 100 K the $\nu_1$-(Ti-O) mode decreases from 629 $cm^{-1}$ and shows a minimum close to 622 $cm^{-1}$ while the $\nu_1$-(Mg-O) mode decreases from approximately 672 $cm^{-1}$ and shows a minimum at 670 $cm^{-1}$, both minima around 250 K. Above 250 K, both modes increase and show a peak at around 350 K. As we can see in Fig. 3, an anomalous behavior was clearly observed between 250 and 350 K in $\nu_1$-(Ti-O) and $\nu_1$-(Mg-O) modes. The lower frequency $\nu_1$-(Nb-O) mode also presents an anomalous behavior between 250 and 350 K but it is almost imperceptive. The anomaly at this mode will be evidenced by differentiate curves subsequently (Fig. 4).

A possible explanation for the observed anomaly can be founded if we consider that in this temperature interval the monoclinic and tetragonal phase coexists. Note that this coexistence was not previously observed when the PMN-PT was studied using high-resolution synchrotron x-ray powder diffraction[11] or Rietveld studies.[22] However, infrared spectroscopy is a punctual technique and as consequence, it is more sensible to discrete phase transitions as the present case. Thus, in this work we proposed that these anomalous behavior for $\nu_1$-(Nb-O), $\nu_1$-(Ti-O) and $\nu_1$-(Mg-O) modes in Fig. 3 can be associated with a fine *monoclinic* → *monoclinic* + *tetragonal* → *tetragonal* phase transition in PMN-PT with 35% $PbTiO_3$ composition, similar to that observed in PZT ceramics.[10] This phase coexistence can be understood based on distortions associated to a unit cell of the monoclinic and tetragonal phase during the phase transition. If the system changes from a low to a high symmetry, a slight shift of the stretching mode to high frequency can be expected. If the associated phase transition is well defined, we can expect a well defined discontinuity, but if this phase transition occurs progressively, typically observed in solid solutions such as PZT,[10] an increasing must be observed in the observed mode as a result from superposition from both low and high frequency modes, as shown in Fig. 3.

Under the point of view of the group representation, the possibilities for proposed phase transition can be attested understanding the possible symmetries involving the $BO_6$ octahedron. The cubic symmetry, point group $O_h$, presents an $F_{1u}$ species and a single triply degenerate band structure is expected, since three equivalent axes exist in the case of the cubic lattice. When the cubic phase transforms into tetragonal phase, point group $C_{4v}$, an $E$ and $A_1$ species appear with cubic triple degeneracy partially removed. Consequently, a double band structure is expected for tetragonal phase. Finally, the $E$ and $A_1$ transforms into $A'$ and $A''$ species when the symmetry change from tetragonal to monoclinic.[23] Thus, a double partially degenerate band structure is also expected for the monoclinic phase. Therefore, the monoclinic and tetragonal phase coexistence in PMN-PT must be interpreted as a superposing of two double partially degenerate band structures.

To show that anomalies observed in $\nu_1$-(Ti-O) and $\nu_1$-(Mg-O) modes are also present for $\nu_1$-(Nb-O) mode, and consequently establish the valid temperature interval that monoclinic and tetragonal phases coexist, we derive each spectrum in Fig. 3 as a function of the temperature ($d\nu/dT$) and plots $d\nu/dT$ versus temperature in Fig. 4. As we can see, all spectra clearly show inflection points, including the low frequency $\nu_1$-(Nb-O) mode, which are associated to behavior changes for each spectra. In this figure, we can observe two distinct inflection points at 198 K and 300 K. The monoclinic phase predominates from low temperature up to 198 K, when the coexistence monoclinic-tetragonal phases occurs between 198 K and 300 K. Above 300 K, the tetragonal phase up to investigated temperature predominates (400 K).

In Fig. 4, the new monoclinic-tetragonal phase coexistence for PMN-PT was added to the new phase diagram recently proposed by Noheda *et al.*[11] near the MPB, considering a vertical line for x = 0.35 % of $PbTiO_3$. This result represents an advance in MPB comprehension of the PMN-PT and complements results previously obtained from different techniques. Probably, the phase coexistence involving a monoclinic phase is a common phenomenon observed in ferroelectric solid solutions that exhibits a MPB region, such as PZT and PMN-PT. In the PZT system, Ragini *et al.*[24] have shown that the monoclinic-tetragonal phases coexist as a result of a first order phase transition between the low temperature monoclinic and high temperature tetragonal phases.

After this study, some questions remain. Why was this monoclinic-tetragonal phase coexistence not previously observed? What are the causes of this phase coexistence? This monoclinic ($M_C$) phase in PMN-PT system was only recently observed (August 2002).[11] More recently (February 2003), a $M_B$ and $M_C$ monoclinic phase evidence were also observed in the MPB region, using dielectric measurements and Rietveld studies,[22] but neither of these works predicts with precision the coexistence of the monoclinic-tetragonal phase in the PMN-PT. Most likely, this monoclinic phase was not previously observed because the tools used are macroscopic techniques. On the other hand, infrared spectroscopy is a punctual technique and for this reason it was possible to detect this coexistence. At this stage, it is important to place emphasis on the fact that each IR measurement was performed in isotherm regime: each spectrum was recorded after one hour of stable temperature. For this reason, we believe that the anomalous behavior observed in this work between 198 K and 300 K really corresponds to monoclinic-tetragonal phase coexistence. Responding to the second question, it is possible that this phase coexistence is a result of monoclinic ($M_C$) phase instabilities at higher temperatures, as previously observed for monoclinic phase ($M_A$) in PZT ceramics[11] but the complete comprehension of this coexistence question will be possible through studies with others PMN-PT compositions in MPB investigated in details.

## Conclusions

In this work, PMN-35PT (0.35 % of $PbTiO_3$) single crystals grown from vertical Bridgman method were studied using spectroscopic infrared studies. Through this study, it was possible to follow the behavior, as a function of the temperature, of the $\nu_1$-*stretching* modes (Nb-O, Ti-O and Mg-O stretch) in $BO_6$ octahedron in $ABO_3$ structure of the PMN-PT. All these frequency modes presented an anomalous behavior between 250 and 350 K. These anomalies were attributed to the monoclinic-tetragonal phase coexistence. The 198-300 K interval temperature of monoclinic-tetragonal validly was determined from derivative dν/dT, leading to *monoclinic* → *monoclinic* + *tetragonal* → *tetragonal* phase transition for this specific PMN-PT composition. Probably, this phase coexistence is a result of the monoclinic phase instabilities at higher temperatures. The complete understanding of this phase coexistence on PMN-PT system remains unclear and will be possible through studies with other PMN-PT compositions near MPB.

**Acknowledgements**

We would like to express our gratitude to Laboratory for Advanced Materials from Stanford University for providing the samples. We would also like to thank Conselho Nacional de Desenvolvimento Científico e Tecnológico (CNPq), Fundação Coordenação de Aperfeiçoamento de Pessoal de Nível Superior (CAPES), Fundação para o Desenvolvimento da Unesp (FUNDUNESP) and Fundação de Amparo à Pesquisa do Estado de São Paulo (FAPESP) for financial support. We also acknowledge MSc. L.H.Z. Pelaio for technical assistance with IR measurements.

## Figure captions

**Fig. 1**: Infrared absorbance spectra of the PMN-35PT crystal recorded for several temperatures in the 100-470 K range.

**Fig. 2**: Infrared absorbance spectrum of the PMN-35PT crystal recorded at 100 K showing the experimental data, lorentzian and fitting curves.

**Fig. 3**: Behavior of the $\nu_1$-*stretching* modes (Nb-O, Ti-O and Mg-O stretch) in $BO_6$ octahedron in $ABO_3$ structure of the PMN-PT as a function of the temperature. Each point represents a result from computational fitting using three lorentzian functions for all temperature ranges and line is solely an eye guide.

**Fig. 4**: Behavior of the $d\nu/dT$ versus temperature (see text). In this figure, line is solely an eye guide. Dot vertical lines indicate the curve inflection point and the limit between *monoclinic* ($M_C$) → *monoclinic* ($M_C$) + *tetragonal* (T) → *tetragonal* (T).

**Fig. 5**: Changes proposed on PMN-PT phase diagram around the MPB were based on results from this work suggesting the following *monoclinic* → *monoclinic* + *tetragonal* → *tetragonal* phase transition. Solid circles were obtained from Noheda *et al.*[11] and open square results from infrared spectroscopy results obtained in the present work.

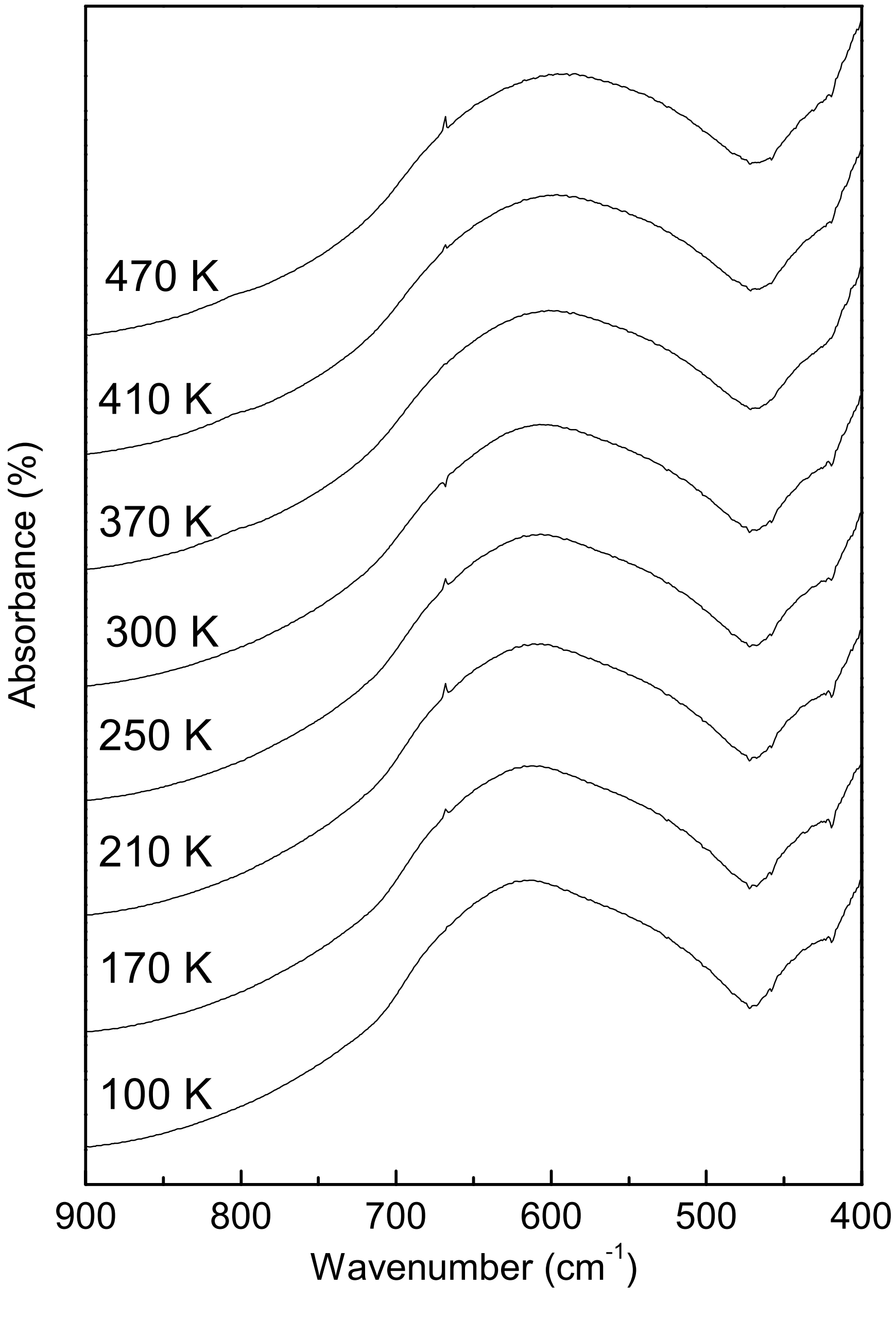

Fig. 1: Araújo et al.

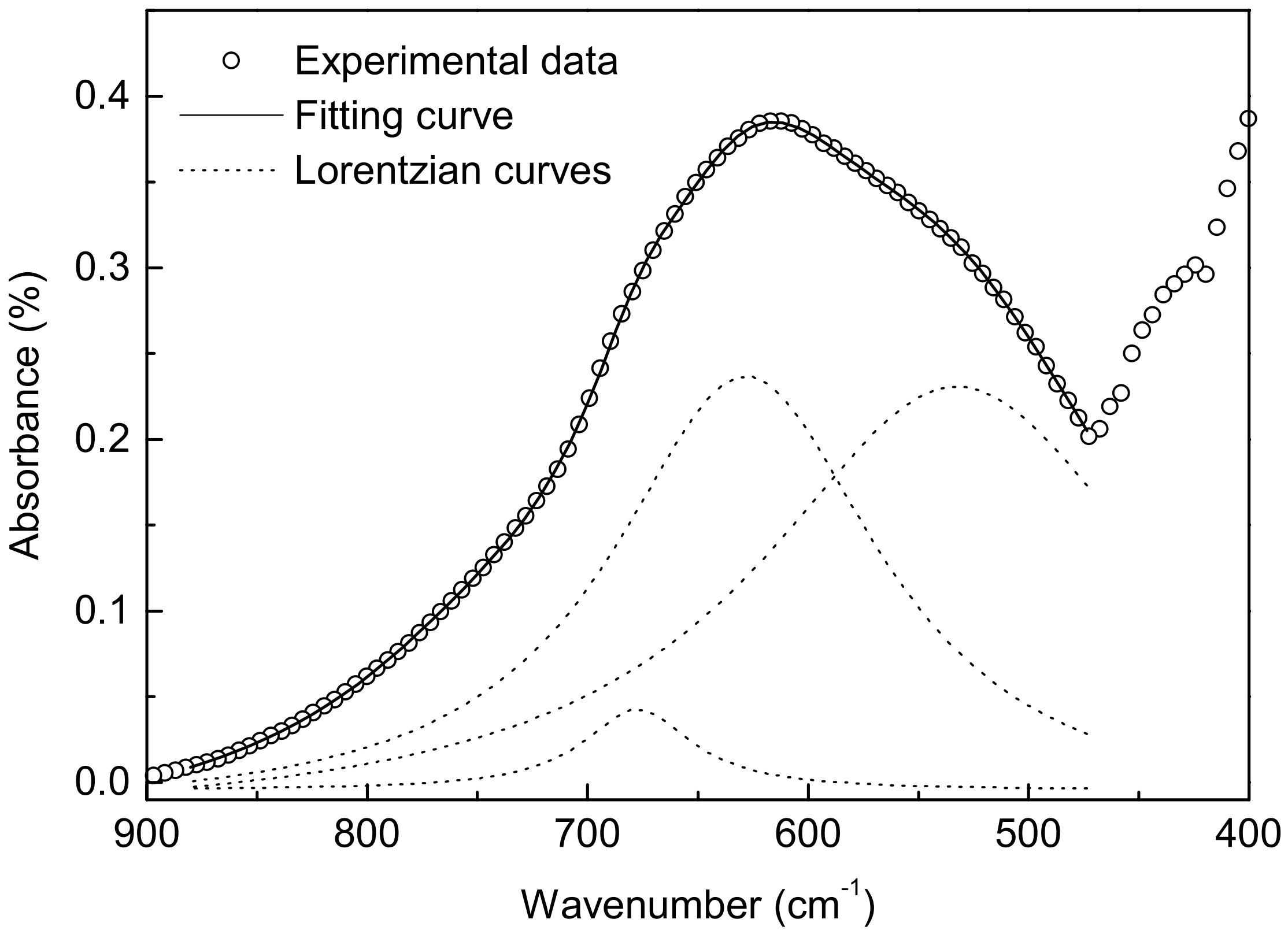

Fig. 2: Araújo et al.

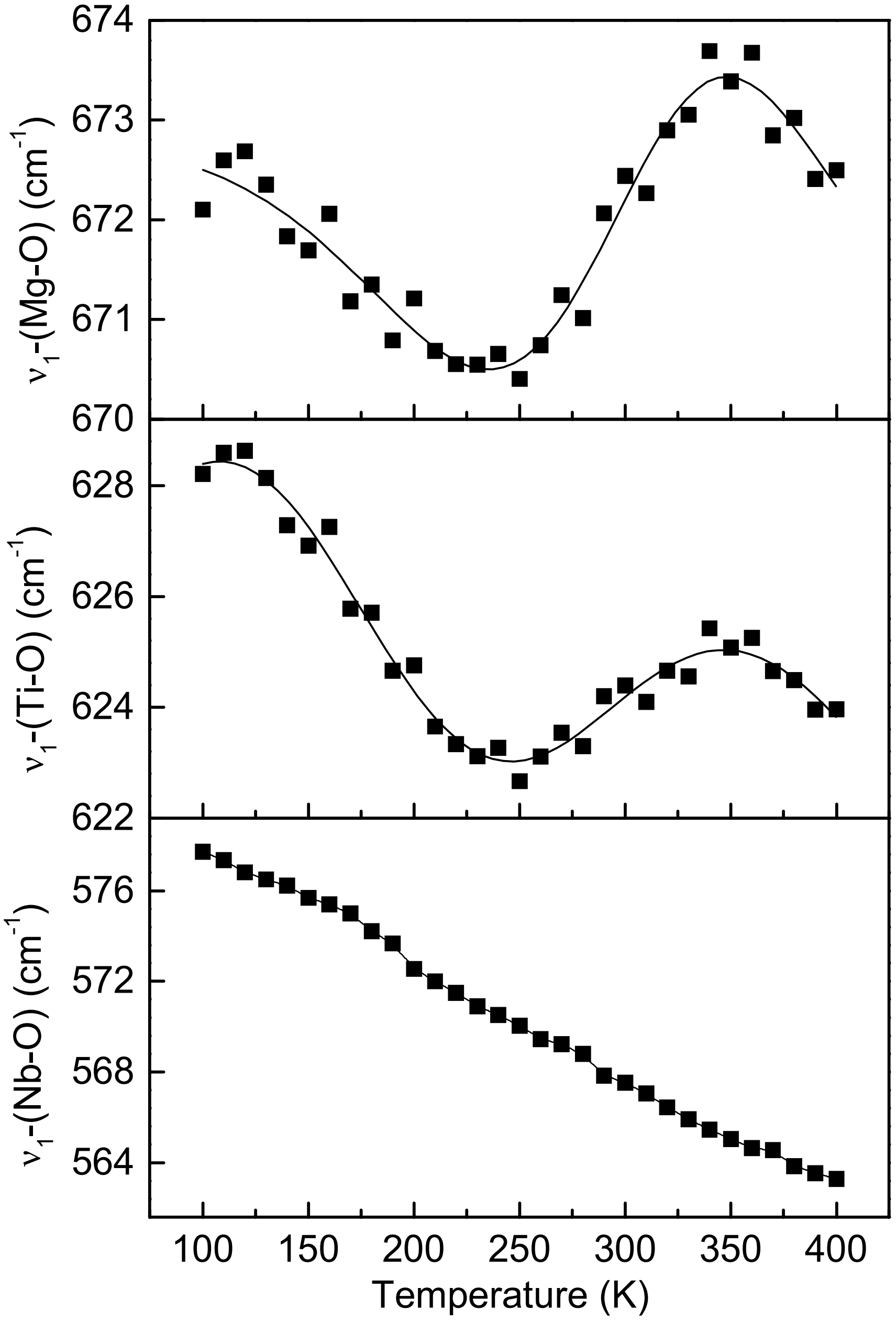

Fig. 3: Araújo et al.

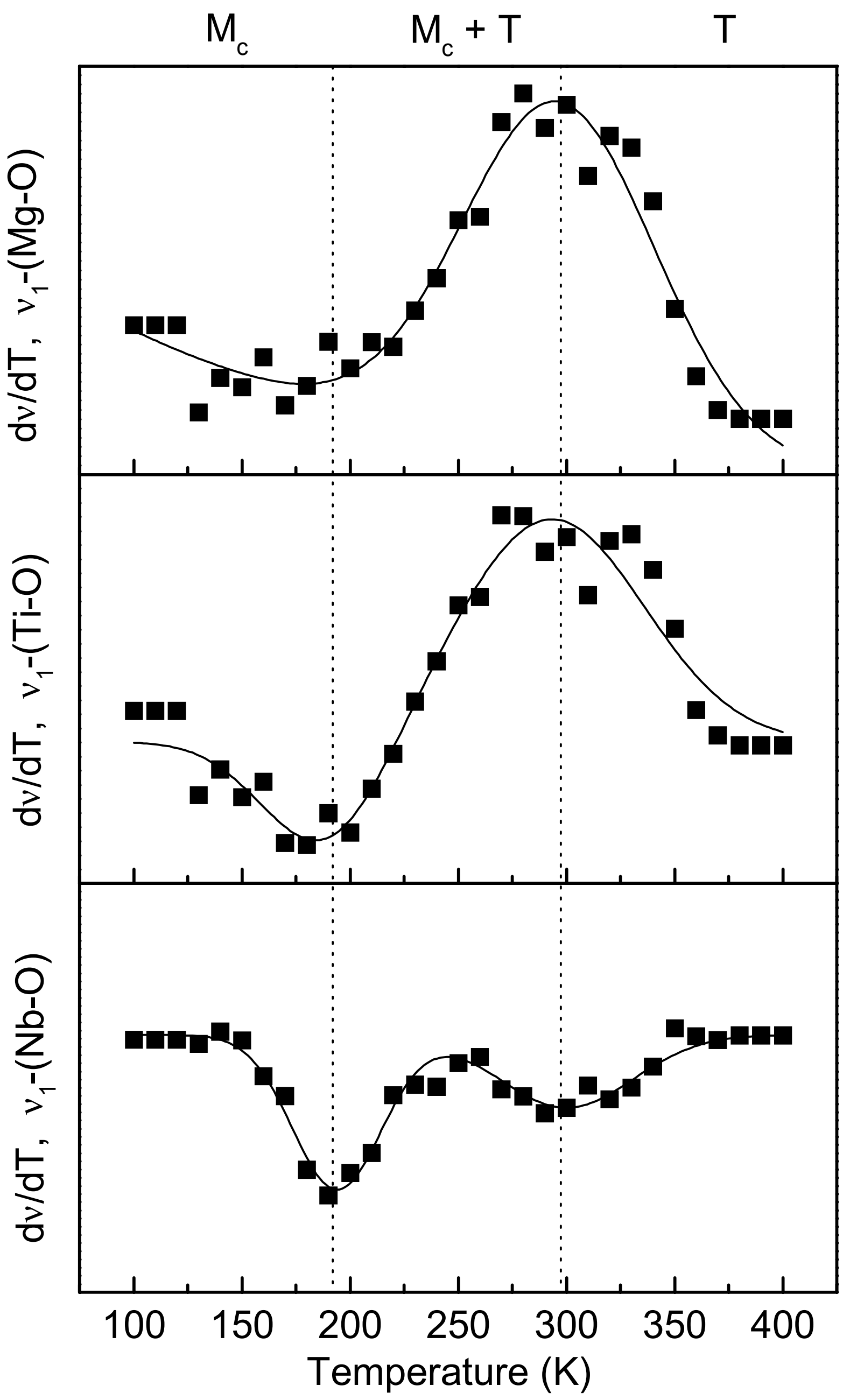

Fig. 4: Araújo et al.

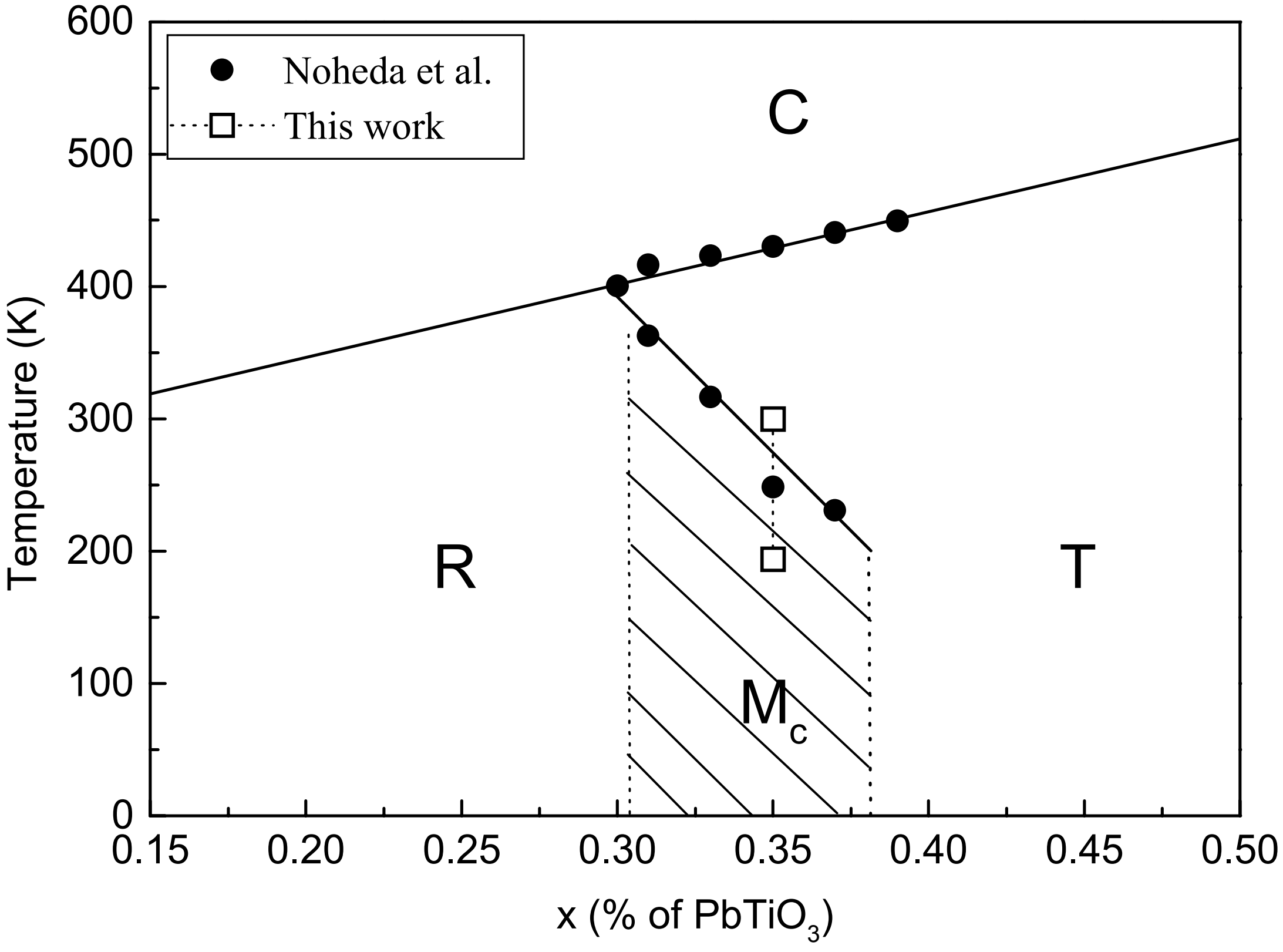

Fig. 5: Araújo et al.